\documentclass[pdftex, 12pt]{article}
\usepackage{newtxtext, newtxmath}
\usepackage{amsmath}
\usepackage[top=2.54cm, bottom=2.54cm, left=2.54cm, right=2.54cm]{geometry}
\usepackage{setspace}
\usepackage{authblk}
\usepackage{afterpage}
\usepackage[hidelinks]{hyperref}
\usepackage[pdftex]{graphicx}
\usepackage{booktabs}
\usepackage{setspace}
\usepackage{titlesec}
\usepackage{tocloft}
\usepackage{parskip}
\usepackage{amsmath}
\usepackage{amsfonts}
\usepackage{lscape}

\usepackage{biblatex}
\usepackage{float}
\usepackage{multirow}
\usepackage{longtable}
\usepackage[table,xcdraw]{xcolor}
\usepackage{caption}
\usepackage{subcaption}
\newcommand{\Rlogo}{\protect\includegraphics[height=1.8ex,keepaspectratio]{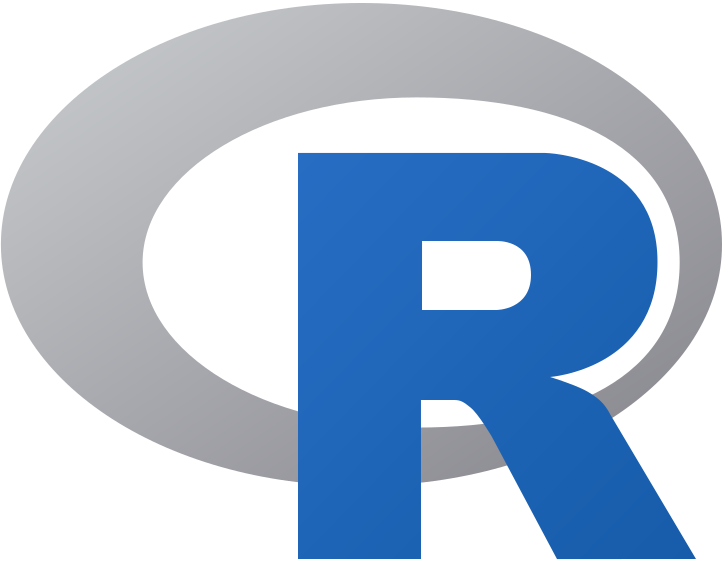}}
\usepackage{titlesec}

\titleformat{\paragraph}
{\normalfont\normalsize\bfseries}{\theparagraph}{1em}{}
\titlespacing*{\paragraph}
{0pt}{3.25ex plus 1ex minus .2ex}{1.5ex plus .2ex}

\title{Parameter estimation of the four-parameter Harris extended Weibull distribution with applications to real-life data}
\author[1]{Prithul Chaturvedi}
\author[1*]{Himanshu Pokhriyal}
\affil[1]{Prasanna School of Public Health, Manipal Academy of Higher Education, Manipal, KA, India \protect\\ \{prithulc, hpokhriyal1908\}@gmail.com}
\affil[*]{Corresponding author} 
\date{}
\begin{document}

\maketitle
\thispagestyle{empty}
\cleardoublepage

\thispagestyle{empty}
\begin{center}
    \LARGE\textbf{Abstract}
\end{center}

This paper explores the extension of the classical two-parameter Weibull distribution to a four-parameter Harris extended Weibull (HEW) distribution. The flexibility of this probability distribution is illustrated by the varying shapes of HEW density function. Estimation of HEW parameters is explored using estimation methods such as the least-squares, maximum product of spacings, and minimum distance method. We provide Bayesian inference on the random parameters of the HEW distribution using Metropolis-Hastings algorithm to sample from the joint posterior distribution. Performance of the estimation methods is assessed using extensive simulations. The applicability of the distribution is demonstrated against three variants of the Weibull distribution on three real-life datasets.

\textbf{Keywords}: Harris extended Weibull distribution; least-squares method; maximum product of spacings method; minimum distance method; Newton-Raphson method; genetic algorithm; Bayesian inference; Metropolis-Hastings algorithm

\pagebreak
\pagenumbering{arabic}
\section{Introduction} \label{intro}

The Weibull distribution is one of the most versatile probability distributions in statistical literature to describe real-world happenings. What once was a tool used to describe structural strength now is a popular in analysis of skewed lifetime data in biology, finance, engineering, and insurance. Improving the flexibility of lifetime distributions has been explored in the past since elastic models are sought-after to model data that are often skewed. This is commonly referred to as an extension of a distribution. 
\newline \newline
Lai et. al. introduced a modified Weibull distribution capable of modelling a bathtub hazard function\cite{1179794}. The three-parameter exponentiated Weibull distribution introduced by Mudholkar and Srivastava is popularly implemented in modelling survival data due to its unimodal, increasing, decreasing, and bathtub hazard functions\cite{mudholkar1993exponentiated}. This distribution also has a two-parameter form in which the product of the two shape parameters is fused. The truncated Weibull distribution is a modified version of said base distribution where truncation in support of the random variable results in heavy-tailed Weibull distribution; popular applications are apparent in finance and insurance industries\cite{kizilersuproperties}. Another such extension of the Weibull distribution is the Marshal-Olkin extended Weibull distribution (MOEW) which belongs to the Marshal-Olkin family of distributions where an additional parameter is introduced to the Weibull distribution\cite{marshall1997new}. 
\newline \newline
Based on the probability generating function of Harris distribution, Aly and Benkherouf worked upon the idea of introducing two additional shape parameters to any baseline distribution, and the newly obtained distribution would fall under the Harris extended (hereby HE) family of distribution\cite{aly2011new}. This technique was picked to become one of the most well-known methods of obtaining better flexibility of probability distributions. A baseline distribution can be any existing probability distribution where the random variable $X > 0$. Over the years, many authors have contributed to expanding the HE family of distribution by proposing baseline distributions such as exponential\cite{doi:10.1080/03610926.2013.851221}.
\newline \newline
In general, if $\theta$ and $k$ are the two new shape parameters, then the HE density function \cite{jose2015} is given by
\begin{equation}
    f_X(x) = \frac{\theta^{1/k}f_0(x)}{[1-\bar{\theta}\bar{F}_0(x)^k]^{(k+1)/k}}; \hspace{1.5mm} x>0; \hspace{1.5mm} \theta, k > 0\label{eq:HE}
\end{equation}
where $\bar{\theta} = 1-\theta$, $f_0(x)$ is the baseline probability density function (pdf), $F_0(x)$ is the baseline cumulative density function (cdf), and $\bar{F_0}(x) = 1-F_0(x)$ is the survival function (sf).
\newline \newline
When the baseline distribution is a two-parameter Weibull distribution, we get the Harris extended Weibull (HEW) distribution which was proposed by Batsidis and Lemonte \cite{batsidis2015harris}, which is a generalization of the MOEW distribution. Both of these distributions have been discussed in Jose et. al. \cite{jose2018} for their application to quality control data.
\newline \newline
Parameter estimation of any distribution is fundamentally the most important aspect of statistical modelling. In this paper, we compare seven different estimation methods to find out the best one to estimate the four parameters of the HEW distribution for different sample sizes. The nominated estimation methods are: 
\begin{enumerate}
    \item Maximum likelihood
    \item Ordinary least-squares
    \item Weighted least-squares
    \item Maximum product of spacings
    \item Minimum distance methods
    \begin{enumerate}
        \item Anderson-Darling
        \item Cram\'er-von Mises
    \end{enumerate}
    \item Bayesian analysis with informative priors $Ga(a, b)$ absolute error loss function. \newline
\end{enumerate}
In section \ref{hew}, we visit the HEW distribution theoretically and discuss its properties. In section \ref{est_meth}, we discuss the estimation methods as listed above in more detail and study them using simulations in section \ref{simul}. Section \ref{data analysis} deals with three real-life datasets where section \ref{fhew} will demonstrate how well the HEW distribution fits them as compared to three other forms of the Weibull distribution and section \ref{bhew} will discuss Bayesian fitting of HEW distribution on the same three datasets. All computations for this paper were performed on \Rlogo\cite{RCloud}.
\section{Methods}
\subsection{The Harris Extended Weibull (HEW) Distribution} \label{hew}

 If $X$ is a random variable that is Weibull distributed with shape $\beta$ and rate $\alpha$, its pdf and sf are given by (\ref{eq:dweib}) and (\ref{eq:pweib}) respectively\cite{johnson1995continuous}
\begin{equation}
    f_X(x) = \alpha \beta x^{\beta-1}  exp {\left(-\alpha x^\beta \right)}; \hspace{1.5mm} x \ge 0; \hspace{1.5mm} \alpha, \beta \ge 0 \label{eq:dweib}
\end{equation}
\begin{equation}
    \bar{F}_X(x) = exp {\left(-\alpha x^\beta \right)}; \hspace{1.5mm} x \ge 0; \hspace{1.5mm} \alpha, \beta \ge 0 \label{eq:pweib}
\end{equation}
\flushleft From (\ref{eq:HE}), (\ref{eq:dweib}) and (\ref{eq:pweib}), we obtain the pdf of $X \sim \text{HEW}(\theta, k, \beta, \alpha)$, given by:
\begin{equation}
            f_X(x) = \frac{\theta^{1/k}\hspace{1mm}\alpha \beta x^{\beta-1} \hspace{1mm} \exp \left(-\alpha x^\beta\right)}{\left[1-\bar{\theta}\hspace{1mm} \exp \left( -k \alpha x^\beta \right) \right]^{(k+1)/k}}; \hspace{1.5mm} x > 0;\hspace{1.5mm} \theta, k, \beta, \alpha > 0; \hspace{1.5mm} \bar{\theta} = 1-\theta \label{eq:hew}
\end{equation}

and the cdf given by:

\begin{equation}
            F_X(x) = \left\{\frac{\theta\hspace{1mm}\exp \left(-k\alpha x^\beta\right)}{1-\bar{\theta}\hspace{1mm} \exp \left( -k \alpha x^\beta \right)}\right\}^{1/k}\label{eq:hew_cdf}
\end{equation}

Here, $\theta>0$, $k>0$ and $\beta>0$ are shape parameters and $\alpha>0$ is the rate parameter. When $\theta=k=1$, the HEW distribution is reduced to the Weibull distribution with shape $\beta$ and rate $\alpha$. When $\theta=k=\beta=1$, it we obtain the exponential distribution with rate $\alpha$. When $k=1$, we obtain the Marshall-Olkin extended Weibull distribution with parameters shape 1 $\theta$, shape 2 $\beta$ and rate $\alpha$. Figure \ref{fig:hew} visualizes the flexibility of the HEW distribution for various combinations of shape parameters at fixed rate $\alpha = 1$.

\begin{figure}[H]
    \centering
    \begin{subfigure}[H]{1\textwidth}
         \centering
         \caption{}
         \includegraphics[width=\textwidth]{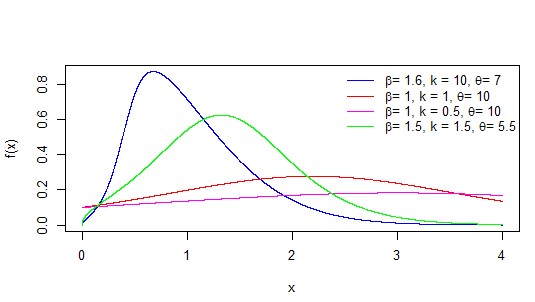}
    \end{subfigure}
    \hfill
     \begin{subfigure}[H]{1\textwidth}
         \centering
         \caption{}
         \includegraphics[width=\textwidth]{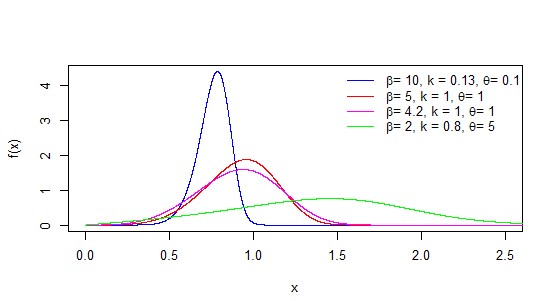}
    \end{subfigure}  
\end{figure}
\begin{figure} \ContinuedFloat
\centering
     \begin{subfigure}[H]{1\textwidth}
         \centering
         \caption{}
         \includegraphics[width=\textwidth]{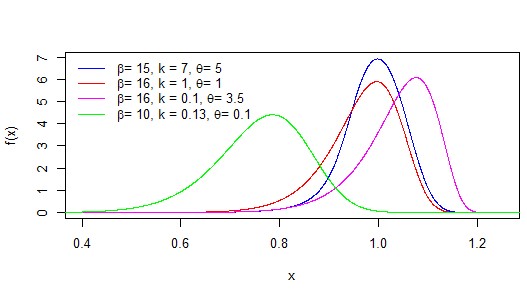}
    \end{subfigure} 
    \hfill
\begin{subfigure}[H]{1\textwidth}
         \centering
         \caption{}
         \includegraphics[width=\textwidth]{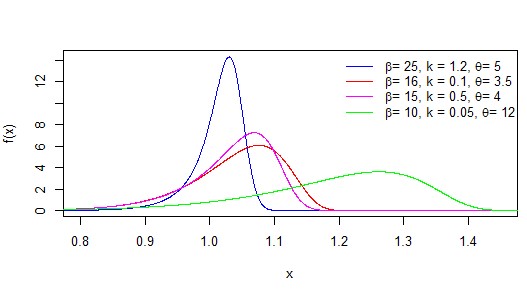}
    \end{subfigure} 
    \caption{Pdf of the HEW distribution at various shape parameters ($\alpha$= 1). This is to demonstrate the distribution's flexibility}
    \label{fig:hew}
\end{figure}

\subsection{Parameter estimation methods}\label{est_meth}
In this section, we discuss 7 different estimation methods\cite{do2015comparison} for the four HEW distribution parameters $\theta$, $k$, $\beta$ and $\alpha$, all considered unknown simultaneously. These methods are considered in the simulation study presented in section \ref{simul}. These parameter estimation methods have an objective function each which can be maximised/minimised either by Newton-Raphson method or Genetic Algorithm w.r.t. the parameters either numerically or by using \Rlogo \hspace{1mm}(\texttt{GA::ga} or \texttt{optim}).

\subsubsection{Maximum likelihood (MLE)}

Let $x_1, x_2, ..., x_n$ be a $n$ observations from the HEW distribution with unknown parameters $\theta$, $k$, $\beta$ and $\alpha >0$. The likelihood and log-likelihood functions are given by (\ref{like}) and (\ref{llike}) respectively. 

\begin{equation}
            L(\theta, k, \beta, \alpha \mid \underline{x}) = \prod_{i=1}^n \frac{\theta^{1/k}\hspace{1mm}\alpha \beta x_i^{\beta-1} \hspace{1mm} \exp \left(-\alpha x_i^\beta\right)}{\left[1-\bar{\theta}\hspace{1mm} \exp \left( -k \alpha x_i^\beta \right) \right]^{(k+1)/k}} \label{like}
\end{equation}

\begin{equation}
        \textit{ln}\hspace{0.5mm}L(\alpha, \beta, k, \theta; \underline{x}) = n\ln \left({ \theta^{1/k}\hspace{1mm} \alpha \hspace{1mm} \beta }\right) + (\beta-1) \sum_{i=1}^n \ln x_i - \alpha \sum_{i=1}^n x_i^\beta - \left({\frac{k+1}{k}}\right) \sum_{i=1}^n \ln{\left( 1-\bar{\theta}\hspace{1mm}e^{-k\alpha x_i^\beta} \right)} \label{llike}    
\end{equation}

The maximum likelihood estimates of the parameters can be obtained by maximising (\ref{llike}).

\subsubsection{Ordinary least-squares (OLS)}

Let $x_{1:n}< x_{2:n}< ...< x_{n:n}$ be the ordered statistics of size $n$ from the HEW distribution with cdf $F_X(x)$ as given by (\ref{eq:hew_cdf}). It is established that
\begin{equation}
    \text{E}[F(x_{i:n})]=\frac{i}{n+1} \hspace{2mm} \text{and} \hspace{2mm}\text{V}[F(x_{i:n})] = \frac{i(n-i+1)}{(n+1)^2(n+2)}
\end{equation}

The OLS estimates of $\theta$, $k$, $\beta$ and $\alpha >0$ are obtained by minimising the following function:

\begin{equation}
    \text{S}(\theta, k, \beta,\alpha \mid \underline{x}) = \sum_{i=i}^n \bigg( F(x_{i:n}\mid \theta, k, \beta,\alpha) - \text{E}[F(x_{i:n})]\bigg)^2
\end{equation}

\subsubsection{Weighted least-squares (WLS)}

Along the same lines as OLS, the WLS estimates are given by minimising the following objective function:

\begin{equation}
    \text{S}(\theta, k, \beta,\alpha \mid \underline{x}) = \sum_{i=i}^n w_i\bigg( F(x_{i:n}\mid \theta, k, \beta,\alpha) - \text{E}[F(x_{i:n})]\bigg)^2
\end{equation}

where $w_i$ is called the correction factor, given by 

\begin{equation}
    w_i=\frac{1}{\text{V}[F(x_{i:n})]}=\frac{(n+1)^2(n+2)}{i(n-i+1)}
\end{equation}

\subsubsection{Maximum product of spacings (MPS)}

Let $x_{1:n}< x_{2:n}< ...< x_{n:n}$ be the ordered statistics of size $n$ from the HEW distribution. Consider the following quantities called uniform spacings of the sample: $D_1=F(x_{1:n}\mid \theta, k, \beta,\alpha)$, $D_{n+1}=1-F(x_{n:n}\mid \theta, k, \beta,\alpha)$, and $D_i=F(x_{n:n}\mid \theta, k, \beta,\alpha)-F(x_{i:n}\mid \theta, k, \beta,\alpha), \hspace{1mm} i=1,2,3,...,n$. Note that here, there are $(n+1)$ spacings of the first order. The MPS estimates of the parameters are the values that maximises the MPS statistic (the geometric mean of the spacings) given by \ref{mps_1}, or it's logarithm $H = \text{log}(G)$.

\begin{equation}
    G(\theta, k, \beta,\alpha \mid \underline{x})=\left( \prod_{i=1}^n D_i\right)^{1/n+1}
    \label{mps_1}
\end{equation}

By considering $0=F(x_{0:n}\mid \theta, k, \beta,\alpha) < 
F(x_{1:n}\mid \theta, k, \beta,\alpha) < ... < F(x_{n:n}\mid \theta, k, \beta,\alpha) < F(x_{(n+1):n}\mid \theta, k, \beta,\alpha)=1$, the log of (\ref{mps_1}) is given by

\begin{equation}
    H(\theta, k, \beta,\alpha \mid \underline{x}) = \frac{1}{n+1} \sum_{i=1}^n \log(D_i)
\end{equation}

The MPS estimation method is as efficient as MLE estimation and the its estimates are consistent under more general conditions then MLE estimates\cite{cheng1983estimating}.

\subsubsection{Minimum distance methods}

In this section, we discuss two estimation procedures based on minimisation of corresponding goodness-of-fit statistics. A common attribute between both these statistics is that they're based on the difference between cdf and empirical cumulative distribution function (ecdf). 

\paragraph{Anderson-Darling (AD)}

Here, we minimise the Anderson-Darling goodness-of-fit statistic, given by (\ref{ad}), to obtain the HEW estimates.

\begin{equation}
    A(\theta, k, \beta,\alpha \mid \underline{x}) = -n-\frac{1}{n} \sum_{i=1}^n (2i-1) \log \bigg( F(x_{i:n}\mid \theta, k, \beta,\alpha)[1-F(x_{(n+1-i):n}\mid \theta, k, \beta,\alpha)] \bigg)
    \label{ad}
\end{equation}

\paragraph{Cram\'er-von Mises (CvM)}

Here, we minimise the Cram\'er-von Mises goodness-of-fit statistic, given by (\ref{cvm}), to obtain the HEW estimates.

\begin{equation}
    C(\theta, k, \beta,\alpha \mid \underline{x}) = \frac{1}{12n} + \sum_{i=1}^n \left( F(x_{i:n}\mid \theta, k, \beta,\alpha) - \frac{2i-1}{2n} \right)^2
    \label{cvm}
\end{equation}

\subsection{Bayesian analysis}

Let $\Theta = \{\theta, k, \beta, \alpha\}$ of the HEW distribution. Let $\pi(\theta), \pi(k), \pi(\beta)$ and $\pi(\alpha)$ be the prior probability distributions (hereby simply priors) of $\theta, k, \beta$ and $\alpha$ respectively. Let $L(\Theta \mid \underline{x})$ represent the joint likelihood function as in (\ref{like}). Hence by Bayes theorem for distributions, the posterior distribution $\pi(\theta, k, \beta, \alpha \mid \underline{x})$ is given by

\begin{equation}
    \pi(\theta, k, \beta, \alpha \mid \underline{x}) \propto \pi(\theta) \times \pi(k) \times \pi(\beta) \times \pi(\alpha) \times L(\Theta \mid \underline{x})
    \label{bayes_1}
\end{equation}

For Bayesian estimation simulations in section \ref{simul}, we simulate data as specified above. We also assume that the four HEW parameters are independently Gamma distributed with corresponding shape and rate hyperparameters. These hyperparameters are estimated based on the maximum likelihood estimation and standard error (from hessian matrix) that act as mean and standard deviation of the Gamma distribution. Metropolis-Hastings (MH) algorithm is used to approximate the required marginal posterior distributions. The proposal distribution chosen is $\text{N}_4(M, \Sigma)$ where  $\Sigma=\begin{bmatrix} 0.01&0\\0&0.01 \end{bmatrix}$.
\section{Results} \label{results}
\subsection{Simulations for parameter estimation} \label{simul}

\begin{figure}[ht]
    \centering
    \includegraphics[scale = 0.6]{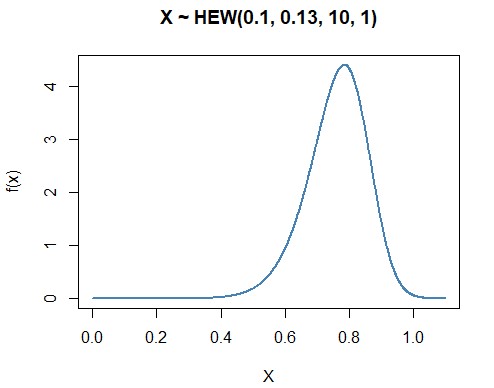}
    \caption{HEW distribution for simulations}
    \label{sim_dens}
\end{figure}

In this section, we examine the comparison of performance of the different estimation methods as discussed briefly in chapter \ref{est_meth}. The sample sizes taken to generate pseudo-random samples from HEW distribution are $n=$ 25, 50, 100, 200. Since the focus of this project is on testing the flexibility of HEW distribution for negatively skewed data, we consider this combination of parameters which yields a negatively skewed density as in Figure \ref{sim_dens}: $\theta=0.1$, $k=0.13$, $\beta=10$, and $\alpha=1$. 

For each sample size, we have generated $10^6$ observations. Root mean-square error (RMSE) and bias was derived to assess the parameters. Time taken per iteration by these estimation methods for all the sample sizes is given in Table \ref{time} and the simulation results are reported in Table \ref{simu_tab}.

\begin{table}[h]
\centering
\resizebox{12cm}{!}{
\begin{tabular}{lllllll}
\hline
\textbf{n}   & \textbf{MLE} & \textbf{OLS} & \textbf{WLS} & \textbf{MPS} & \textbf{AD} & \textbf{CvM} \\ \hline \hline
\textit{10}  & 0.0425       & 0.1435       & 0.1399       & 0.1778       & 0.1900      & 0.1420     \\
\textit{25}  & 0.0460       & 0.2196       & 0.2076       & 0.3207       & 0.3328      & 0.2168    \\
\textit{50}  & 0.0514       & 0.3482       & 0.3248       & 0.5639       & 0.5773      & 0.3402    \\
\textit{100} & 0.0623       & 0.6094       & 0.5612       & 1.0392       & 1.0571      & 0.5873    \\ \hline
\end{tabular}}
\caption{Time taken (seconds) per iteration}
\label{time}
\end{table}

\begin{table}[h]
\centering
\resizebox{\columnwidth}{!}{%
\begin{tabular}{@{}llllllllll@{}}
\toprule
\multirow{2}{*}{\textbf{Method}} &
  \multirow{2}{*}{\textit{\textbf{n}}} &
  \multicolumn{2}{c}{\textit{\textbf{$\theta$}}} &
  \multicolumn{2}{c}{\textit{\textbf{$k$}}} &
  \multicolumn{2}{c}{\textit{\textbf{$\beta$}}} &
  \multicolumn{2}{c}{\textit{\textbf{$\alpha$}}} \\ \cmidrule(l){3-10} 
 &
   &
  \textbf{RMSE} &
  \textbf{Bias} &
  \textbf{RMSE} &
  \textbf{Bias} &
  \textbf{RMSE} &
  \textbf{Bias} &
  \textbf{RMSE} &
  \textbf{Bias} \\ \midrule
\multirow{4}{*}{\textbf{MLE}}   & \textit{25}  & 0.0294 & -0.0017 & 0.0866 & 0.0211  & 1.7826 & -0.6028 & 0.2739 & -0.1243 \\
                                & \textit{50}  & 0.0259 & -0.0072 & 0.0819 & 0.0128  & 1.1258 & -0.2327 & 0.2378 & -0.1293 \\
                                & \textit{100} & 0.0244 & -0.0089 & 0.0695 & 0.0130  & 0.8401 & -0.1180 & 0.2230 & -0.1053 \\
                                & \textit{200} & 0.0218 & -0.0074 & 0.0652 & -0.0039 & 0.6371 & -0.0842 & 0.2211 & -0.0944 \\ \midrule
\multirow{4}{*}{\textbf{OLS}}   & \textit{25}  & 0.0412 & 0.0142  & 0.0806 & 0.0072  & 2.0288 & 0.0011  & 0.3588 & 0.1101  \\
                                & \textit{50}  & 0.0380 & 0.0157  & 0.0780 & 0.0076  & 1.4265 & 0.0204  & 0.3470 & 0.1516  \\
                                & \textit{100} & 0.0364 & 0.0176  & 0.0742 & 0.0069  & 1.0173 & 0.0257  & 0.3411 & 0.1744  \\
                                & \textit{200} & 0.0216 & 0.0085  & 0.0535 & 0.0132  & 0.6575 & 0.0602  & 0.2149 & 0.1047  \\ \midrule
\multirow{4}{*}{\textbf{WLS}}   & \textit{25}  & 0.0419 & 0.0185  & 0.0592 & 0.0120  & 4.2587 & 0.2057  & 0.3604 & 0.0443  \\
                                & \textit{50}  & 0.0416 & 0.0195  & 0.0589 & 0.0128  & 4.1357 & 0.1374  & 0.3551 & 0.0595  \\
                                & \textit{100} & 0.0411 & 0.0197  & 0.0583 & 0.0156  & 3.9244 & -0.0384 & 0.3399 & 0.0783  \\
                                & \textit{200} & 0.0265 & 0.0132  & 0.0347 & 0.0169  & 2.7481 & 0.0034  & 0.2125 & 0.0361  \\ \midrule
\multirow{4}{*}{\textbf{MPS}}   & \textit{25}  & 0.0396 & -0.0066 & 0.1024 & -0.0354 & 2.6001 & 1.6124  & 0.4262 & 0.2761  \\
                                & \textit{50}  & 0.0351 & 0.0064  & 0.0981 & -0.0290 & 1.5686 & 0.8139  & 0.4031 & 0.2479  \\
                                & \textit{100} & 0.0344 & 0.0134  & 0.0916 & -0.0257 & 1.0244 & 0.3917  & 0.3736 & 0.2200  \\
                                & \textit{200} & 0.0202 & 0.0046  & 0.0691 & -0.0138 & 0.6680 & 0.2225  & 0.2125 & 0.0899  \\ \midrule
\multirow{4}{*}{\textbf{AD}}    & \textit{25}  & 0.0282 & 0.0035  & 0.0682 & 0.0052  & 1.6813 & 0.1878  & 0.2631 & 0.0657  \\
                                & \textit{50}  & 0.0239 & 0.0051  & 0.0631 & 0.0056  & 1.1899 & 0.1241  & 0.2341 & 0.0799  \\
                                & \textit{100} & 0.0217 & 0.0064  & 0.0581 & 0.0040  & 0.8576 & 0.0669  & 0.2086 & 0.0796  \\
                                & \textit{200} & 0.0213 & 0.0060  & 0.0557 & 0.0084  & 0.6252 & 0.0599  & 0.2098 & 0.0753  \\ \midrule
\multirow{4}{*}{\textbf{CvM}}   & \textit{25}  & 0.0318 & 0.0002  & 0.0643 & 0.0046  & 2.0299 & 0.5628  & 0.2978 & 0.1177  \\
                                & \textit{50}  & 0.0263 & 0.0035  & 0.0607 & 0.0068  & 1.3895 & 0.3304  & 0.2614 & 0.1155  \\
                                & \textit{100} & 0.0236 & 0.0067  & 0.0565 & 0.0081  & 0.9730 & 0.1871  & 0.2372 & 0.1158  \\
                                & \textit{200} & 0.0217 & 0.0081  & 0.0561 & 0.0110  & 0.6912 & 0.1277  & 0.2305 & 0.1177  \\ \midrule
\multirow{4}{*}{\textbf{Bayes}} & \textit{25}  & 0.0345 & 0.0010  & 0.0846 & -0.0168 & 1.7952 & 0.7020  & 0.2637 & 0.1109  \\
                                & \textit{50}  & 0.0323 & 0.0048  & 0.0783 & -0.0122 & 1.1507 & 0.2420  & 0.2364 & 0.0958  \\
                                & \textit{100} & 0.0303 & 0.0082  & 0.0718 & -0.0111 & 0.8845 & 0.2003  & 0.2413 & 0.1197  \\
                                & \textit{200} & 0.0266 & 0.0092  & 0.0665 & -0.0002 & 0.6023 & 0.1090  & 0.2093 & 0.1012  \\ \bottomrule
\end{tabular}%
}
\caption{Simulation results}
\label{simu_tab}
\end{table}

\subsection{Fitting distributions to real-life data} \label{data analysis}

In this section we discuss data analysis of three datasets from real-life setup to investigate applicability of HEW distribution in model fitting. Brief description and summary statistics of the datasets are given as following and in table \ref{data_sum} respectively.

\begin{enumerate}
    \item \textbf{Bladder Cancer}: uncensored data corresponding to remission times (in months) of a random sample of 128 bladder cancer patients\cite{jose2015}
    \item \textbf{Carcinoma}: survival time of the patients (in months) of a lung cancer clinical trial being conducted on 194 patients with squamous cell carcinoma by the Eastern Cooperative Oncology Group\cite{lagakos1978covariate}
    \item \textbf{Carbon}: 63 observations on breaking stress of carbon fibers (in Gba)\cite{jose2018}
\end{enumerate}

\begin{table}[h]
\centering
\resizebox{15cm}{!}{%
\begin{tabular}{@{}lllllllll@{}}
\toprule
\textbf{Dataset}   & \textbf{N}   & \textbf{Min}  & \textbf{Q1}    & \textbf{Med.} & \textbf{Mean}  & \textbf{Q3}     & \textbf{Max}   & \textbf{Skew.}  \\ \midrule
\textbf{Bladder Cancer}    & 127 & 0.08 & 3.34 & 6.25 & 8.82 & 11.72 & 46.12 & 2.08   \\
\textbf{Carcinoma} & 194 & 1.00    & 8.00     & 14.00   & 18.81 & 24.75  & 101   & 2.078  \\
\textbf{Carbon}    & 63  & 0.39 & 2.09 & 2.85 & 2.74 & 3.28  & 4.90   & -0.198 \\ \bottomrule
\end{tabular}%
}
\caption{Summary statistics of the considered datasets}
\label{data_sum}
\end{table}

\subsubsection{Frequentist HEW fit} \label{fhew}

In this section, we compare the AIC and three goodness-of-fit test scores: Kolmogorov-Smirnov (KS), Anderson-Darling (AD), Cram\'er-von Mises (CvM) of HEW distribution to that of three other variants of the baseline distribution:
\begin{enumerate}
    \item Weibull distribution $Weib(\beta, \alpha)$
    \item Truncated Weibull distribution $tWeib(\beta, \alpha)$ with truncation point $\gamma$ set to the maximum observation value in respective univariate dataset.
    \item Exponentiated Weibull distribution $expWeib(\theta, \alpha)$
\end{enumerate}

The cdf of these distributions are given in Table \ref{cdf_tab}.

\begin{table}[h]
\centering
\resizebox{10cm}{!}{%
\begin{tabular}{@{}lll@{}}
\toprule
\textbf{Distribution}  & \textbf{CDF} & \textbf{Parameters}\\ 
\midrule
\textbf{$\text{Weib}(\beta, \alpha)$} & $1-\exp(-\alpha x^{\beta})$ &  $\alpha, \beta > 0$\\
\textbf{$\text{tWeib}(\beta, \alpha)$} & $\frac{1-\exp(-\alpha x^\beta)}{1-\exp(-\alpha \gamma^\beta)}$ &$\alpha, \beta > 0, \hspace{2mm}\gamma = \text{max}(\underline{x})$\\
\textbf{$\text{expWeib}(\theta, \alpha)$}     & $(1-\exp(-x^{\theta}))^{\alpha}$   &  $\theta, \alpha > 0$                     \\ \bottomrule
\end{tabular}%
}
\caption{CDF of Weib, tWeib, and expWeib distributions}
\label{cdf_tab}
\end{table}

The intuitive reason behind choosing the above-mentioned distributions for comparison is to evaluate whether the two new shape parameters from HEW distribution actually improve flexibility. It is also to contrast the versatility of HEW distribution with other forms of recognized Weibull distribution in literature. We estimate all the parameters of the four distributions under scrutiny by the method of maximum likelihood estimation. The estimates and statistics are given in table \ref{freq_est} and figure \ref{freq_fit} visually shows the fitted HEW distributions on the datasets. From table \ref{freq_est}, one can observe that the smallest AIC statistics are that of the HEW distribution for Bladder Cancer and Carcinoma datasets, which clearly indicates that HEW distribution fits the best to the two datasets among all the considered lifetime distributions. It is also seen that for the Carbon dataset, the truncated Weibull distribution fits the best, but however looking at maximized log likelihood values L of all the distributions, HEW distribution obtains the best results. This shall be further discussed in section \ref{disc}.

\begin{landscape}
\begin{table}[]
\begin{tabular}{llllllll}
\hline
\textbf{Dataset}                                   & \textbf{Distribution}               & \textbf{Estimates}                                          & \textbf{L}                             & \textbf{AIC}                           & \textbf{KS (p-val)}                        & \textbf{AD (p-val)}                       & \textbf{CvM (p-val)}                       \\ \hline \hline
                                                   & {\textit{HEW}} & {\textit{(8.46, 4.79, 0.79,    0.27)}} & {\textit{-399.7}} & {\textit{807.37}} & {\textit{0.03 (1)}}     & {\textit{0.08 (1)}}    & {\textit{0.012 (0.99)}} \\ 
                                                   & Weib                             & (1.12, 9.22)                                                & -402.07                                & 808.14                                 & 0.06 (0.69)                                  & 0.6 (0.64)                                  & 0.1 (0.6)                                    \\  
                                                   & tWeib                            & (1.12, 9.3)                                                 & -401.79                                & 807.59                                 & 0.07 (0.63)                                  & 0.64 (0.61)                                 & 0.11 (0.55)                                  \\ 
\multirow{-4}{*}{\textit{\textbf{Bladder Cancer}}} & expWeib                          & (0.47, 6.42)                                                & -403.13                                & 810.25                                 & 0.05 (0.85)                                  & 0.53 (0.71)                                 & 0.08 (0.67)                                  \\ \hline
                                                   & {\textit{HEW}} & {\textit{(9.05, 4.31, 0.84, 0.13)}}    & {\textit{-753.1}} & {\textit{1514.2}} & {\textit{0.052 (0.68)}} & {\textit{0.31 (0.93)}} & {\textit{0.047 (0.89)}} \\ 
                                                   & Weib                             & (1.24, 20.25)                                               & -756.24                                & 1516.49                                & 0.07 (0.32)                                  & 0.1 (0.6)                                   & 0.99 (0.36)                                  \\ 
                                                   & tWeib                            & (1.23, 20.32)                                               & -756.11                                & 1516.22                                & 0.069 (0.31)                                 & 1.05 (0.33)                                 & 0.17 (0.32)                                  \\ 
\multirow{-4}{*}{\textit{\textbf{Carcinoma}}}      & expWeib                          & (0.42, 12.43)                                               & -757.54                                 & 1519.07                                & 0.11 (0.03)                                  & 1.56 (0.16)                                 & 0.27 (0.17)                                  \\ \hline
                                                   & {\textit{HEW}} & {\textit{(14.23, 1.00, 1.80, 0.43)}}   & {\textit{-83.07}} & {\textit{174.14}} & {\textit{0.07 (0.89)}}  & {\textit{0.34 (0.91)}} & {\textit{0.05 (0.86)}}  \\ 
                                                   & Weib                             & (3.29, 3.04)                                                & -84.69                                 & 173.38                                 & 0.09 (0.66)                                  & 0.64 (0.61)                                 & 0.11 (0.52)                                  \\ 
                                                   & tWeib                            & (3.18, 3.07)                                                & -84.22                                  & 172.45                                 & 0.1 (0.62)                                   & 0.67 (0.58)                                 & 0.12 (0.5)                                   \\ 
\multirow{-4}{*}{\textit{\textbf{Carbon}}}         & expWeib                          & (1.01, 8.51)                                                & -93.77                                 & 191.54                                 & 0.17 (0.05)                                  & 2.41 (0.06)                                 & 0.46 (0.05)                                  \\ \hline
\end{tabular}
\caption{Parameter estimates, AIC and goodness-of-fit statistics}
\label{freq_est}
\end{table}
\end{landscape}

\begin{figure}[H]
    \centering
    \begin{subfigure}[H]{0.65\textwidth}
         \centering
         \caption{}
         \includegraphics[width=\textwidth]{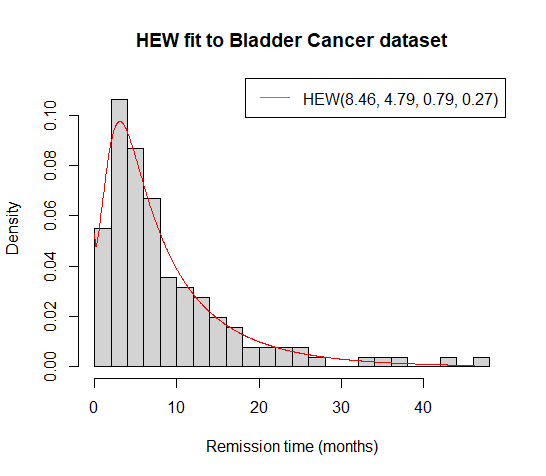}
    \end{subfigure}
     \hfill
     \begin{subfigure}[H]{0.65\textwidth}
         \centering
         \caption{}
         \includegraphics[width=\textwidth]{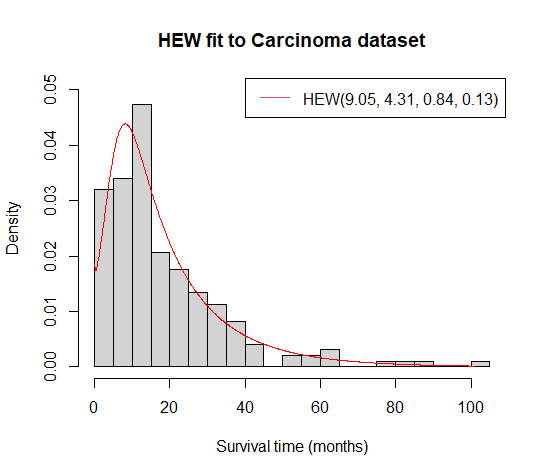}
    \end{subfigure}  
\end{figure}
\begin{figure}\ContinuedFloat
\centering
     \begin{subfigure}[H]{0.65\textwidth}
         \centering
         \caption{}
         \includegraphics[width=\textwidth]{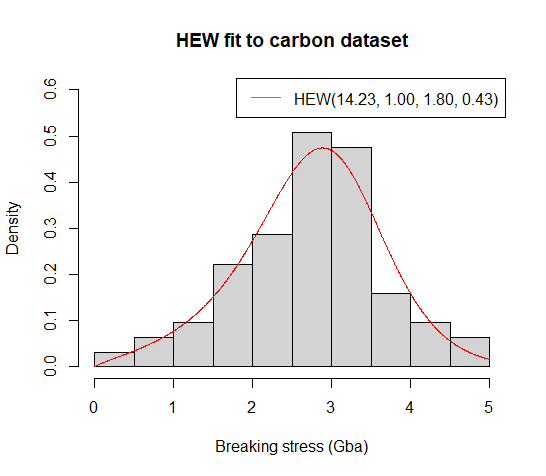}
    \end{subfigure}    
    \caption{HEW distribution fitted to real-life datasets}
    \label{freq_fit}
\end{figure}

\subsubsection{Bayesian HEW fit} \label{bhew}

\paragraph{Setup}\label{setup}
\textbf{Choosing priors and prior elicitation} \label{prior}

Since the HEW parameters are all $>0$, we can consider any lifetime distribution to define priors. In this paper, we choose the two-parameter gamma distribution, i.e., $\theta$, $k$, $\beta$, and $\alpha$ $\sim$ \textit{Ga(shape, rate)}. This is a suitable distribution to proceed with since its flexibility is well-credited in literature and it has always been one of the most popular priors for rate and shape parameters.

Since we have three datasets under analysis, all coming from three domains of expertise, it is ideally expected from researchers to fetch information by means of extensive research and consulting field experts. Due to time constraints and limitation of resources, we proceed to use the frequentist HEW estimates and corresponding standard error to elicit required gamma hyperparameters. The estimated hyperparameters are given in Table \ref{gamma}.

\begin{table}[h]
\centering
\resizebox{15cm}{!}{%
\begin{tabular}{lllllllll}
\hline
\multirow{2}{*}{\textbf{Dataset}} & \multicolumn{2}{c}{\textbf{$\theta$}}               & \multicolumn{2}{c}{\textbf{$k$}}                 & \multicolumn{2}{c}{\textbf{$\beta$}}                 & \multicolumn{2}{c}{\textbf{$\alpha$}}                 \\ \cline{2-9} 
                                  & \textit{\textbf{Shape}} & \textit{\textbf{Rate}} & \textit{\textbf{Shape}} & \textit{\textbf{Rate}} & \textit{\textbf{Shape}} & \textit{\textbf{Rate}} & \textit{\textbf{Shape}} & \textit{\textbf{Rate}} \\ \hline
\textbf{Bladder Cancer}           & 0.66                    & 0.08                   & 2.14                    & 0.45                   & 12.81                   & 16.28                  & 1.45                    & 5.39                   \\
\textbf{Carcinoma}                & 0.97                    & 0.13                   & 3.20                    & 0.69                   & 19.00                   & 21.65                  & 1.27                    & 11.57                  \\
\textbf{Carbon}                   & 0.21                    & 0.01                   & 1.98                    & 2.00                   & 3.42                    & 1.93                   & 0.26                    & 0.56                   \\ \hline
\end{tabular}%
}
\caption{Gamma hyperparameters}
\label{gamma}
\end{table}

\textbf{Likelihood function} \label{likelihood}

The likelihood and log-likelihood functions of \textbf{$n$} random samples from HEW distribution is given by (\ref{like}) and (\ref{llike}) respectively. These mathematical forms are not used anywhere in the analysis since all estimations were rendered using the HEW pdf on \Rlogo.

\textbf{Loss function}\label{loss}

We choose the absolute error loss function, under which the posterior median is found to be the Bayes estimate. This decision is taken upon assessment of the nature of the parameter distributions under analysis.

\paragraph{Bayesian analysis}\label{posterior}

Using MH algorithm, we obtain the estimates under the above-specified loss function. Table \ref{bvsf_tab} shows the computed frequentist (in subsection \ref{fhew}) and Bayes estimates along with statistics to compare the goodness-of-fit on the datasets and figure \ref{bvsf_fig} visually shows the same contrast. It is evident that the added prior information, although basically obtained by frequentist HEW estimates, have an effect on the Bayes estimates and hence they differ by a small margin from the frequentist ones. The BIC and KS goodness-of-fit test suggest that the frequentist estimates fit better to all the three datasets.

\begin{table}
\centering
\resizebox{15cm}{!}{%
\begin{tabular}{llllll}
\hline
\textbf{Dataset}                          & \textbf{Method}                             & \textbf{Estimates}                               & \textbf{L}                     & \textbf{BIC}                   & \textbf{KS (p-val)}               \\ \hline
                                          & { \textit{Frequentist}} & { (8.46, 4.79, 0.79, 0.27)}  & { -399.70} & { 818.78}  & { 0.03 (1)}     \\
\multirow{-2}{*}{\textbf{Bladder Cancer}} & { \textit{Bayesian}}    & { (8.51, 4.76, 0.77, 0.29)}  & { -399.87} & { 819.12}  & { 0.05 (0.95)}  \\ \hline
                                          & { \textit{Frequentist}} & { (9.05, 4.31, 0.84, 0.13)}  & { -753.10} & { 1527.28} & { 0.052 (0.68)} \\
\multirow{-2}{*}{\textbf{Carcinoma}}      & { \textit{Bayesian}}    & { (7.62, 4.62, 0.88, 0.11)}  & { -753.11} & { 1527.29} & { 0.057 (0.57)} \\ \hline
                                          & { \textit{Frequentist}} & { (14.23, 1.00, 1.80, 0.43)} & { -83.07}  & { 182.71}  & { 0.07 (0.89)}  \\
\multirow{-2}{*}{\textbf{Carbon}}         & { \textit{Bayesian}}    & { (14.17, 1.01, 1.81, 0.46)} & { -84.00}  & { 184.57}  & { 0.12 (0.31)}  \\ \hline
\end{tabular}%
}
\caption{Frequentist vs Bayesian estimates}
\label{bvsf_tab}
\end{table}

\begin{figure}[ht]
    \centering
    \begin{subfigure}[H]{0.7\textwidth}
         \centering
         \caption{}
         \includegraphics[width=\textwidth]{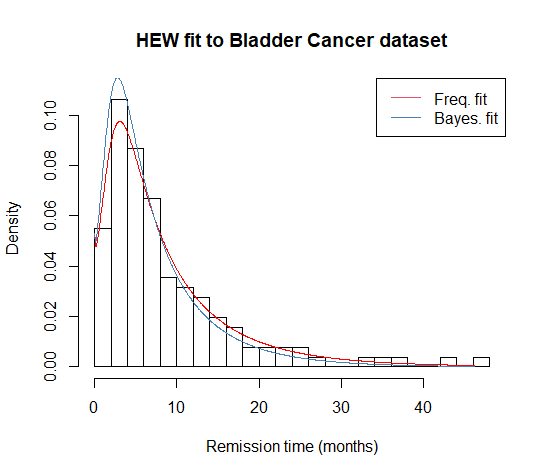}
    \end{subfigure}
\end{figure}
\begin{figure} \ContinuedFloat
     \begin{subfigure}[H]{0.7\textwidth}
         \centering
         \caption{}
         \includegraphics[width=\textwidth]{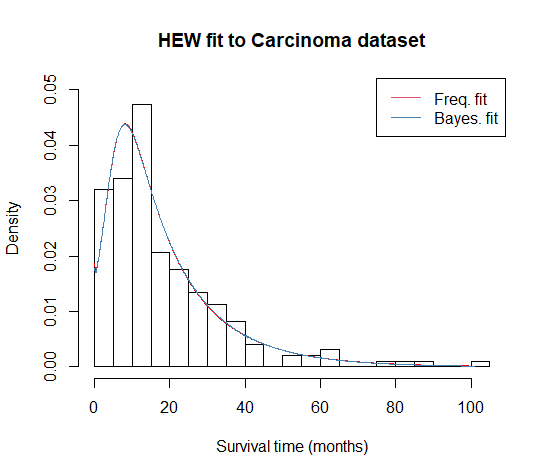}
    \end{subfigure}  
\hfill
\centering
     \begin{subfigure}[H]{0.7\textwidth}
         \centering
         \caption{}
         \includegraphics[width=\textwidth]{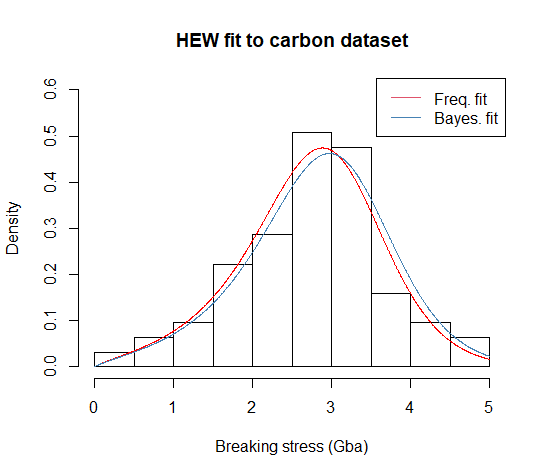}
    \end{subfigure}    
    \caption{Frequentist vs Bayesian fit}
    \label{bvsf_fig}
\end{figure}

Table \ref{width} reports the estimates obtained by both approaches, corresponding confidence/credible intervals and and their respective widths. If $\delta$ is the frequentist estimate of the parameter $\Delta$ and $\sigma/\sqrt{n}$ is it's standard error, the asymptotic confidence interval (ACI) of $\Delta$ at $(1-\alpha)$ confidence level is given by

\begin{equation}
    \label{aci}
    \delta \pm 1.96 \frac{\sigma}{\sqrt{n}}
\end{equation}

The highest posterior density (HPD) interval for the Bayes estimate $\lambda$, which is the shortest interval among all of the credible intervals, is given by the following definition for unimodal distributions:

If $f(.)$ denotes the pdf and $F(.)$ denotes the cdf of the (random) parameter $\lambda$, then the HPD interval $(a, b)$ of $\lambda$ at $(1-\alpha)$ confidence level is obtained by minimizing the function

\begin{equation}
    \label{hpdi}
    g(a,b)=\{F(b)-F(a)-(1-\alpha)\}^2 + k\{f(b)-f(a)\}^2
\end{equation}

where $k>0$ is a tuning parameter that ensures that both the terms are zeroed. In this project, the empirical HPD intervals were obtained using the \Rlogo \hspace{1mm} function \texttt{TeachingDemos::emp.hpd}.
\vspace{10mm}
\begin{table}[h]
\centering
\resizebox{16cm}{!}{%
\begin{tabular}{lclllllll}
\hline
\multirow{2}{*}{\textbf{Dataset}} &
  \multirow{2}{*}{\textbf{Parameter}} &
  \multicolumn{3}{c}{\textit{\textbf{Frequentist}}} &
   &
  \multicolumn{3}{c}{\textit{\textbf{Bayesian}}} \\ \cline{3-5} \cline{7-9} 
 &
   &
  \textbf{F. Est.} &
  \textbf{ACI} &
  \textbf{Width} &
   &
  \textbf{B. Est.} &
  \textbf{HPDI} &
  \textbf{Width} \\ \hline
\multirow{4}{*}{Bladder Cancer} & $\theta$ & 8.46  & (6.69, 10.24) & 3.54  &  & 8.51  & (8.45, 8.50)   & 0.05 \\
                                & $k$      & 4.79  & (4.24, 5.35)  & 1.11  &  & 4.76  & (4.74, 4.78)   & 0.04 \\
                                & $\beta$  & 0.79  & (0.76, 0.83)  & 0.07  &  & 0.77  & (0.76, 0.86)   & 0.10 \\
                                & $\alpha$ & 0.27  & (0.24, 0.31)  & 0.08  &  & 0.29  & (0.28, 0.30)   & 0.02 \\ \hline
\multirow{4}{*}{Carcinoma}      & $\theta$ & 9.05  & (7.98, 10.13) & 2.14  &  & 7.62  & (7.61, 7.64)   & 0.03 \\
                                & $k$      & 4.31  & (3.98, 4.65)  & 0.68  &  & 4.62  & (4.61, 4.63)   & 0.02 \\
                                & $\beta$  & 0.84  & (0.82, 0.87)  & 0.05  &  & 0.88  & (0.86, 0.89)   & 0.03 \\
                                & $\alpha$ & 0.13  & (0.12, 0.15)  & 0.03  &  & 0.11  & (0.08, 0.12)   & 0.04 \\ \hline
\multirow{4}{*}{Carbon}         & $\theta$ & 14.23 & (6.83, 21.64) & 14.80 &  & 14.17 & (14.15, 14.19) & 0.04 \\
                                & $k$      & 1     & (0.83, 1.18)  & 0.35  &  & 1.01  & (1.01, 1.03)   & 0.02 \\
                                & $\beta$  & 1.8   & (1.57, 2.04)  & 0.47  &  & 1.81  & (1.76, 1.86)   & 0.10 \\
                                & $\alpha$ & 0.43  & (0.23, 0.64)  & 0.41  &  & 0.46  & (0.43, 0.5)    & 0.07 \\ \hline
\end{tabular}%
}
\caption{Frequentist vs Bayesian: Estimates, Intervals and Width}
\label{width}
\end{table}
\section{Discussion} \label{disc}

This paper is an attempt to study the suitability of HEW distribution in detail where we discussed the distribution's purpose, importance, properties and need in statistical modeling. Numerous estimation methods were discussed and a simulation study on was carried to find out the best-suited method for HEW distribution. Out of the 6 frequentist and 1 Bayesian methods, the Anderson-Darling test statistic's minimum distance method yielded the best results in general i.e., consistently low RMSE and closest-to-zero bias in majority of the cases, across all sample sizes. The RMSE were observed to consistently decrease with increase in sample size for all the parameters. Hence, one may consider resorting to AD estimation for the best HEW distribution estimates. The ML estimation method estimates faster than all the other methods by significant margins across all sample sizes.

We demonstrated HEW distribution's applicability to three datasets from real-life setups and compared its performance with three other versatile distributions that are popular in literature. The HEW distribution outperforms the other 3 probability distributions in terms of goodness-of-fit tests but falls short to \textit{tWeibull} distribution for Carbon dataset on AIC scores. Akaike Information Criteria penalizes models on complexity and hence a four-parameter distribution could not do better than a two-parameter distribution even though it obtained the smaller maximized likelihood value (L). Two popular approaches in estimation- frequentist and Bayesian were adopted in the above-said data analysis and compared. The classical approach (under ML estimation) yielded better results under no foundational prior information on the parameters from experts.\linebreak

\textbf{Conflict of interest statement}: The authors declare no conflict of interest.

\pagebreak

\printbibliography

\end{document}